\documentclass[conference]{IEEEtran}

\usepackage{cite}      

\usepackage{graphicx}  

\usepackage{psfrag}    

\usepackage{subfigure} 

\usepackage{url}       

\usepackage{stfloats}  

\usepackage{amsmath}   
\usepackage{multirow}
\usepackage{array}
\begin{document}

\title{A Low Complexity Spectrum Sensing Scheme For Estimating Frequency Band Edges In Multi-standard Military Communication Receivers}

\author{\authorblockN{S. J. Darak, A.P. Vinod}
\authorblockA{Universit\'{e} Europ\'{e}enne de Bretagne (UEB) and Sup\'{e}lec \\
Nanyang Technological University\\
Nanyang Avenue, Singapore - 639798\\
Email: {dara0003,asvinod}@ntu.edu.sg}
\and
\authorblockN{E. M-K. Lai}
\authorblockA{Institute of Engineering and Adv. Technology\\
Massey University\\
Albany, New Zealand\\
Email: E.Lai@massey.ac.nz}}


%


\maketitle

\begin{abstract}
In a typical multi-standard military communication receiver, fast and reliable spectrum sensing unit is required to extract the information of multiple channels (frequency bands) present in a wideband input signal. In this paper, an energy detector based on our reconfigurable filter bank, in \cite{darak06}, for detecting the edge frequencies of the channels is proposed. Simulation results are presented to show the trade-off between the time required to calculate edge frequencies of all the channels and the maximum normalized error in estimating the edge frequencies. The proposed method is compared with existing energy detector methods for complexity and performance. It is shown that, for a fixed number of input samples, error decreases with time in the proposed algorithm as compared to other methods where error is constant. Design examples and simulations show that the complexity of the proposed method is lower than the other methods for a given error in estimating the edge frequencies.
\end{abstract}

%

\section{Introduction}
In a multi-standard military communication receiver (MMCR), it is required to sense the wideband input signal consisting of multiple channels (frequency bands) of distinct bandwidths (BWs) and unknown center frequencies. When compared to multi-standard commercial wireless communication receivers such as cognitive radios (CR) \cite{Haykin02}, MMCR has only little priory information about the input signal. The key task of MMCR is to sense the entire spectrum as fast as possible in order to accurately detect the frequency edges of multiple channels present in a wideband input signal. Due to the channel fading, interference and hidden terminal problems, a reliable, low complexity and fast spectrum sensing is one of the most vital requirements in MMCRs.

    There are various filter bank based spectrum sensing approaches in the literature such as single-stage sensing using matched filtering \cite{Cabric03}, energy based detection \cite{Cabric03}, cyclostationary feature based detection (CFD) \cite{Cabric03} or more accurate two-stage sensing techniques such as coarse energy detection followed by serial energy detection \cite{Luo04} or energy detection followed by CFD \cite{Maleki05}. Matched filtering is the optimum method for signal detection since it maximizes the signal to noise ratio. But match filtering requires prior knowledge of the input signal which  makes it non-suitable for MMWRs. Energy detection is the most common method for signal detection and easy to implement. But it has poor performance at low signal to noise ratio (SNR). The CFD takes the advantage of cyclostationary properties of the input signal and have better performance than energy detection at low SNR. But the computation of spectral correlation function in CFD is a highly complex task. Hence, to meet the complexity, time and reliability requirements of MMCRs, two-stage spectrum sensing has been proposed in \cite{Luo04, Maleki05}. It consists of coarse detection stage followed by the fine detection stage. For example in \cite{Luo04, Maleki05}, energy detection is performed in the coarse detection stage where detection time is more important. Then, if necessary, either energy detection \cite{Luo04} or CFD \cite{Maleki05} is performed in the second stage. The advantage of implementing CFD in second stage is the complexity reduction because CFD is performed only when energy detection fails to identify the signal.

This paper deals with the networks where spectrum is not sparse i.e. spectrum occupancy is well above 20$\%$ and hence the entire wideband spectrum needs to be searched simultaneously to detect the channels present in the input. As shown in Fig. 1, the wideband input signal consists of multiple channels (Channels 1, 2 and 3) of distinct BWs with arbitrary edge frequencies (${f_{r1}}$, ${f_{f1}}$), (${f_{r2}}$, ${f_{f2}}$) and (${f_{r3}}$, ${f_{f3}}$) respectively. In MMCRs, the channel BWs and their edge frequencies vary over time as shown in Fig. 1 and these two parameters are often not governed by a fixed rule as opposed to commercial communication receivers. Hence, accurate estimation of edge frequencies of multiple channels present in the wideband input spectrum is a key task in MMCRs.

\begin{figure}[!h]
\centering
\includegraphics[width=3in]{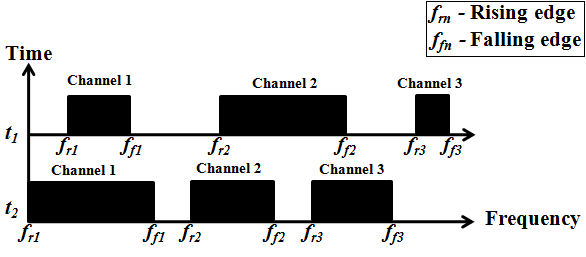}
\caption{Time varying wideband input signal in MMCR.}
\end{figure}

We have proposed the reconfigurable and low complexity filter bank (FB) in \cite{darak06} for channelization operation in MMCRs. The proposed FB provides fine control over the subband BW. In this paper, we focus on the coarse detection part of the two-stage spectrum sensing which detects multiple channels using an energy detector using the reconfigurable filter bank in \cite{darak06} and present an algorithm to accurately find the edges of the channels present in the wideband input spectrum. The proposed method does not require the calculation of the power spectral density using fast Fourier transform or wavelet transform for detecting the edges of the channels.

The rest of the paper is organized as follows. The proposed filter bank based energy detector is introduced in Section II followed  by  detailed discussion on  the  edge detection algorithm in Section III. Simulation results and complexity analysis are given in Section IV. Section V has our conclusion.

\section{Proposed Filter bank based energy detector}

For the analysis, following assumptions about the input signal spectrum are made:

1) The number of channels, their BWs and locations are unknown to the MMCR. This information may change over time but assumed to be unchanged during one cycle of sensing.

2) The range of frequencies over which sensing needs to be performed (bandwidth of input signal) is known to MMCR.

3) The power spectral density within each channel is assumed to be almost flat.

4) The input noise is addictive white Gaussian noise (AWGN) with zero mean and unit variance.

The block diagram of the proposed scheme is shown in Fig. 2. It consists of the (\emph{M+1})-band filter bank, proposed in \cite{darak06}, followed by an energy detector block for each subband. Outputs of each energy detector block are fed to the edge detection algorithm which calculates the edge frequencies of all the channels present in the input.

\begin{figure}[!h]
\centering
\includegraphics[width=3in]{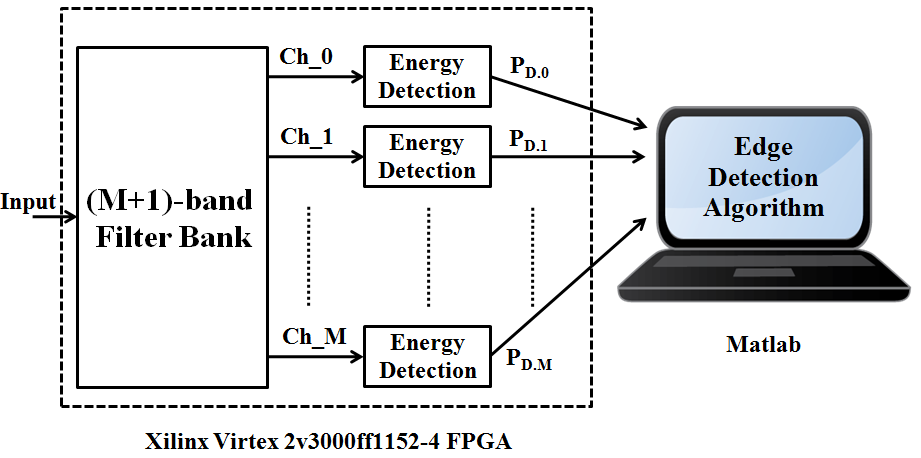}
\caption{Proposed spectrum sensing scheme.}
\end{figure}

\subsection{Filter Bank}
In \cite{darak06}, we have proposed a low complexity and recon- figurable FB based channelizer (Channelization is the process of extraction of individual radio channels (frequency bands) from the wideband input signal) for MMCRs. The FB is based on coefficient decimation, interpolation and frequency masking techniques. We have employed the same FB for spectrum sensing approach in this work. Re-using the same FB for channelization as well as spectrum sensing results in complexity reduction and this advantage has been studied in \cite{mahesh07}.

The block diagram of the FB is shown in Fig. 3. It consists of two stages: 1) a low pass linear phase FIR filter
called the modal filter, ${H_a(Z^{M/D})}$, with CD-II, interpolation and complementary filter, ${H_c(Z^{M/D})}$, to obtain multiband response and 2) bank of masking filters to extract desired channels of interest. The first stage of the FB provides distinct bandwidth channels. The number of channels depends on interpolation factor \emph{M} and channel bandwidth depends on decimation factor \emph{D} and interpolation factor \emph{M}. In \cite{darak06}, it is shown that by changing the decimation factor, \emph{D}, the subband BW of the FB can be changed. The outputs of modal filter and complementary filter are fed to respective fixed masking filter banks. Each masking filter will extract the desired channel for which the masking filter is designed, by masking the frequencies of other channels.

In this paper, the same design example of 9-channel FB (i.e. \emph{M}=8) with integer values of \emph{D} ranging from 3 to 7, as discussed in \cite{darak06}, is considered. The ${F_{pass}}$ and ${F_{stop}}$ of modal filter are selected as 0.1 and 0.115 respectively. For \emph{D}=5, all the subbands are of uniform BW. The operation of the proposed FB for channelization and spectrum sensing differs only in the way \emph{D} is changed. In channelization, \emph{D} is changed depending upon the channel BW while for spectrum sensing operation, \emph{D} spans through all the values in one cycle of spectrum sensing irrespective of the channel BW.

\begin{figure}[!h]
\centering
\includegraphics[width=3in]{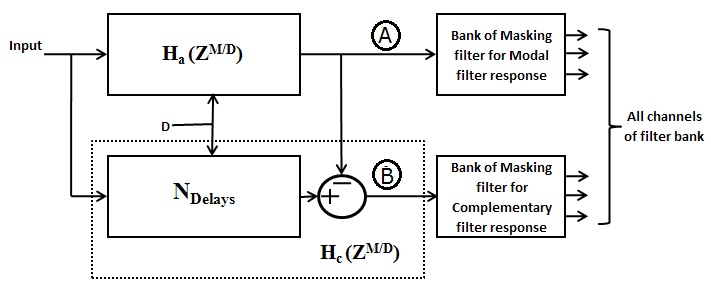}
\caption{Block diagram of reconfigurable filter bank.}
\end{figure}

\subsection{Energy Detection}
The aim of the energy detector block is to calculate the energy of each subbands. Consider the received input signal, $y(n)$ as
\begin{eqnarray}
y(n)= s(n) + w(n).															
\end{eqnarray}

where $s(n)$ is the primary user signal and $w(n)$ is the AWGN. For energy detection method, we compute the decision metric as
\begin{eqnarray}
P= \sum_{n=0}^{N}|y(n)|^{2}.															
\end{eqnarray}

The decision metric, \emph{P}, is compared with a threshold, ${T_o}$, to determine the presence of a primary signal. The threshold, ${T_o}$, is determined  based on the noise level in the  signal and stopband attenuation of the FB. 

In the proposed method, $(M+1)$ energy detector blocks determine the decision metric, \emph{P}, for each value of \emph{D} and pass it to the edge detection algorithm. The matrix \emph{P} has $(M+1)$ columns and one row for each value of \emph{D}. Thus, for the FB discussed above, matrix \emph{P} has 9 columns and 5 rows. The comparison of ${P_{D.M}}$ ( where ${P_{D.M}}$ corresponds to the decision metric for band \emph{M} and decimation factor \emph{D}) with ${T_o}$ is done in the edge detection algorithm stage.

\subsection{One cycle of spectrum sensing}
One cycle of spectrum sensing corresponds to the maximum time required to scan through the entire frequency range with the required frequency resolution. In the proposed method, one cycle of spectrum sensing consists of:

1) First step is to set \emph{D} to its middle value. For the design example considered here, \emph{D} takes integer values from 3 to 7 and hence \emph{D} is set to 5.

2) Run the simulation to obtain \emph{N} samples at the input of each energy detector block and calculate decision metric, ${P_{D.M}}$, using equation (2) for each of the (\emph{M+1}) i.e. 9 subbands.

3) Run the edge detection algorithm to calculate the edge frequencies of all channels present in the input.

4) Repeat step 2 for next higher and lower values of \emph{D} and then go to step 3 i.e. for \emph{D} =4 and \emph{D}=6 and run the edge detection algorithm using ${P_{4.M}}$, ${P_{5.M}}$ and ${P_{6.M}}$.

5) Repeat step 4 for all values of \emph{D}.

\section{Edge detection algorithm}
In this section, our edge detection algorithm to accurately find the edge frequencies of channels is presented. The input to the algorithm is a matrix, \emph{P}. The effect of \emph{D} on the subband BW is shown in Fig. 4 using different colours. Fig. 4(a) shows the modal filter response where subband BW decreases with \emph{D} and Fig. 4(b) shows the complementary filter response where subband BW decreases with increase in \emph{D}.

The first step in edge detection algorithm is to divide the entire frequency range into smaller bands and calculate the energies, ${E_{0}}$, ${E_{1}}$,.... present in these bands. This is done serially from band-0 to band-9. For example, ${E_{0}}$ is equal to ${P_{3.0}}$, ${E_{1}}$ is equal to difference between ${P_{4.0}}$ and ${P_{3.0}}$ and so on. Here we have assumed that ${P_{D.M}}$ is available for all \emph{D}=3,4,5,6,7. If ${P_{D.M}}$ is available only for \emph{D}=4,5,6, then ${E_{0}}$ is equal to ${P_{4.0}}$, ${E_{1}}$ is equal to difference between ${P_{5.0}}$ and ${P_{4.0}}$ and so on. Once, ${E_{0}}$ to ${E_{4}}$ are calculated, then next values are easily obtained using simple subtraction. For example, ${E_{6}}$ is obtained by subtracting ${E_{4}}$ and ${P_{7.1}}$ from ${P_{6.1}}$. Also, note that ${E_{0}}$, ${E_{5}}$, ${E_{10}}$,... are equal to the respective ${P_{D.M}}$ values i.e. ${P_{3.0}}$, ${P_{7.1}}$ and ${P_{3.2}}$ respectively.

\begin{figure}[!h]
\centering
\includegraphics[width=3in]{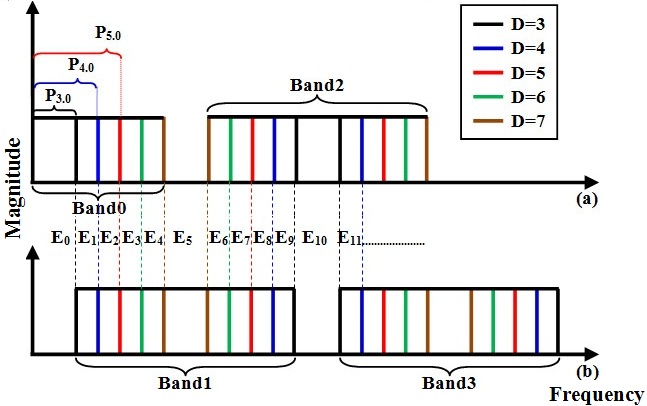}
\caption{Frequency response of filter bank (a) Modal filter response (b) Complementary filter response.}
\end{figure}

The next step is to find whether the primary signal is present into these bands or not. This is done by comparing the band energies, ${E_{x}}$, with the threshold, ${T_o}$. If the energy is greater than ${T_o}$, then signal is present. Otherwise, signal is not present. Then, by using the serial search algorithm, bands containing the edges of the channels are identified based on the comparison between the energy present in that band and its adjacent bands. For example, if ${E_{0}}$, ${E_{1}}$ are less than ${T_o}$ and ${E_{2}}$ is greater than ${T_o}$, then it is concluded that the rising edge is present in the band with energy, ${E_{2}}$. Similarly, if ${E_{5}}$ is less than ${T_o}$ and ${E_{4}}$ is greater than ${T_o}$, then it is concluded that the falling edge is present in the band with energy, ${E_{4}}$. In this way, bands containing either rising or falling edges of the channels are identified using the proposed method.

After identifying the bands containing the edges, next step is to find the approximate edge frequency, ${f_{approx}}$. The simplest approach is to set the edge frequency equal to the centre frequency of the band in which edge is identified. In this approach, the maximum error in calculating the edge frequency is half of the BW. In order to reduce the error further, we are calculating the edge frequency using prediction method which is based on the comparison between the energies present in that band and its adjacent band. This is in accordance with our assumption that power spectral density in each channel is almost flat. For example, if ${E_{2}}$ is much less than ${E_{3}}$, then there are more chances that the rising edge is present at the end of the band rather than at the centre of the band. Similarly, if ${E_{2}}$ and ${E_{3}}$ are almost same, then there are more chances that the rising edge is present at the start of the band rather than at the centre of the band. This helps in further improving the accuracy of edge detection algorithm.

\section{Simulation Results And Complexity Analysis}
In this section, simulation results of the proposed algorithm are presented. The proposed scheme is tested under two different spectral scenarios where wideband input signal consists of multiple channels whose specifications are given in Table I. Note that all the frequencies mentioned in this paper are normalized with respect to the sampling frequency, $f_{samp}$. The 9-channel FB with \emph{D}= 3, 4, 5, 6, 7 is considered here. The modal filter of the FB has passband and stopband frequencies as 0.1 and 0.115 respectively with -30 dB stopband attenuation. The objective is to find the edge frequencies of all the channels present in input. The edge  frequencies obtained using our algorithm, $f_{approx}$, and corresponding error with respect to actual edge frequencies, $f_{actual}$, are given in Table I. The percentage error in the edge frequency is calculated using (3). Note that the errors are low.


\begin{eqnarray}
Error= \frac{|(f_{actual} - f_{approx})|\times 2}{f_{samp}} \times 100.															
\end{eqnarray}

\begin{table}[!h]
\renewcommand{\arraystretch}{1.3}
\caption{Edge frequecies}
\label{table:1}
\begin{center}
\begin{tabular}{|c|c|c|c|c|c|}
\hline
$\textbf{Input}$ & $\textbf{Frequency}$ & $\textbf{Channel}$ & $\boldsymbol {f_{actual}}$ & $\boldsymbol {f_{approx}}$ & $\textbf{Error}$ \\

 & $\textbf{bands}$ & $\textbf{frequency}$ &  &  &  \\

\hline
\multirow{6}{*}{Input 1} &
\multirow {2}{*}{Channel 1} & $f_{rising}$ & 0  & 0  & 0 $\%$\\
\cline{3-6}
                           &  & $f_{falling}$ & 0.13  & 0.128  & 0.2 $\%$ \\
\cline{2-6}

& \multirow {2}{*}{Channel 2} & $f_{rising}$ & 0.3  & 0.299  & 0.1 $\%$  \\
\cline{3-6}
                           & & $f_{falling}$ & 0.65  & 0.642  & 0.8 $\%$ \\

\cline{2-6}
& \multirow {2}{*}{Channel 3} & $f_{rising}$ & 0.78  & 0.781  & 0.1 $\%$ \\
\cline{3-6}
                            & & $f_{falling}$ & 0.89  & 0.881  & 0.9 $\%$ \\
\hline

\multirow{8}{*}{Input 2} &
\multirow {2}{*}{Channel 1} & $f_{rising}$ & 0.06  & 0.058  & 0.2 $\%$ \\
\cline{3-6}
                           &  & $f_{falling}$ & 0.16  & 0.165  & 0.5 $\%$ \\
\cline{2-6}

& \multirow {2}{*}{Channel 2} & $f_{rising}$ & 0.34  & 0.339  & 0.5 $\%$ \\
\cline{3-6}
                           & & $f_{falling}$ & 0.49  & 0.5  & 1$\%$ \\

\cline{2-6}
& \multirow {2}{*}{Channel 3} & $f_{rising}$ & 0.65  & 0.663  & 1.3 $\%$ \\
\cline{3-6}
                            & & $f_{falling}$ & 0.77  & 0.76  & 1 $\%$ \\
\cline{2-6}
& \multirow {2}{*}{Channel 4} & $f_{rising}$ & 0.89  & 0.892  & 0.2 $\%$ \\
\cline{3-6}
                            & & $f_{falling}$ & 1  & 1  & 0 $\%$\\

\hline

\end{tabular}
\end{center}
\end{table}

\subsection{Error Vs Complexity}
For  the  FB architecture  considered  above,  we  compare the maximum error in edge frequency calculation for the proposed method, tree-structured quadrature mirror FB based energy detector (TQMFB) \cite{nar1_08}, DFTFB based energy detector and tree structure DFTFB based energy detector (TDFTFB) \cite{nar2_09}. Since the multiplication is the most complex and power consuming operation in a digital FB, the comparison results in terms of number of real-valued multipliers are shown in Table II. It can be observed that the proposed method offers the best tradeoff between error and multiplication complexity. Note that the error reduction caused by prediction algorithm is not considered here. The relation between error and number of multipliers for different approaches, when different number of channels are present in the FBs, is plotted in Fig. 5. It can be observed that, the proposed FB requires fewer number of multipliers (and hence less computational complexity) than the other methods for a given error.

\begin{table}[!h]
\renewcommand{\arraystretch}{1.3}
\caption{Complexity comparison}
\label{Complexity comparison}
\begin{center}
\begin{tabular}{|c|c|c|}
\hline
$\textbf{Approaches}$ & $\textbf{Max. Error}$ & $\textbf{No. of Multipliers}$ \\
\hline
Proposed method (8-band) & 1.875 $\%$ & 303 \\
\hline
TQMFB \cite{nar1_08} (8-band) & 4.75 $\%$ & 770 \\
\hline
DFTFB (8-band) & 4.75 $\%$ & 328 \\
\hline
TDFTFB \cite{nar2_09} (16-band) & 1.6 $\%$ & 800 \\
\hline
\end{tabular}
\end{center}
\end{table}

\begin{figure}[!h]
\centering
\includegraphics[width=3in]{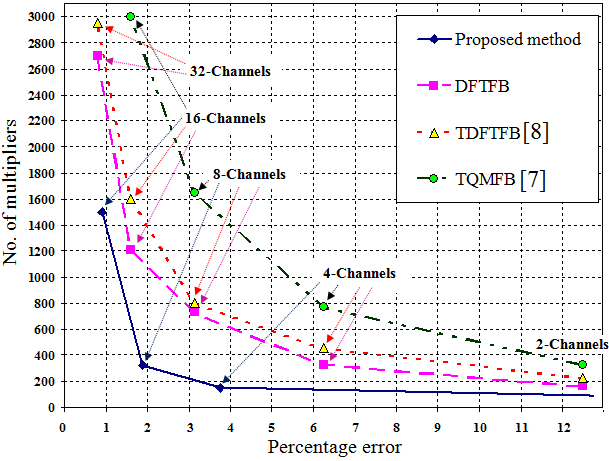}
\caption{Percentage error Vs Complexity comparison.}
\end{figure}

\subsection{Error Vs Time}
The dependence of performance (error and time) on different values of \emph{D} for the proposed method and DFTFB based energy detection is shown in Fig. 6. The number of samples at the input of energy detector is kept constant. It can be observed that the proposed method requires slightly longer detection time than the DFTFB. This is because the proposed FB has slightly higher group delay than other FBs \cite{darak06}. However, due to the reconfigurable architecture of the proposed FB, error decreases with time as ${P_{D.M}}$  is available for more values of \emph{D} as shown in Fig. 6. While in other methods, FB needs to be re-designed to further reduce the error. Note that the error reduction caused by the prediction algorithm is not considered here. The results of TQMFB are similar to DFTFB and hence are not discussed here.

\begin{figure}[!h]
\centering
\includegraphics[width=3in]{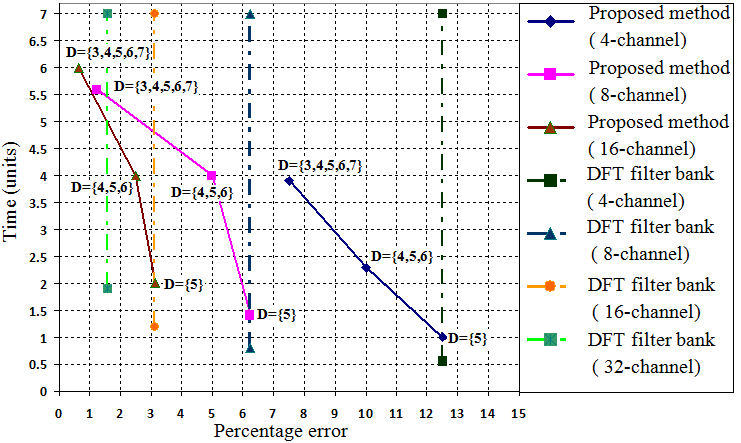}
\caption{Percentage error Vs Time comparison.}
\end{figure}

\section{Conclusion}
We have proposed a filter bank  based energy  detector for obtaining the accurate edge frequencies of the channels present in wideband input signal. The  proposed method does not require calculation of power spectral density using fast Fourier transform or wavelet transform for detecting the edges of the channels. Simulation results shows that the proposed method is computationally more efficient than the other methods for a given error in edge frequency. The proposed method requires slightly more sensing time than the other  methods due to higher group delay. However, due to reconfigurability property of the proposed method, error can be further reduced using the same architecture as compared to other methods where filter bank needs to be re-designed to further reduce the error.





%

\end{document}